\documentclass[aps,prd,showpacs,nobibnotes,twocolumn,
nobalancelastpage,preprintnumbers,
nofootinbib]{revtex4-1}
\usepackage{graphicx}
\usepackage{amsmath,amssymb}
\DeclareMathAlphabet{\mathpzc}{OT1}{pzc}{m}{it}
\usepackage{epstopdf}
\usepackage{topcapt}

\begin{document}

\title{Closing the Window on Strongly Interacting Dark Matter}

\author{Gregory D. Mack}

\author{Adithya Manohar}
\affiliation{Department of Physics and Astronomy,
 Ohio Wesleyan University,
Delaware, Ohio 43015\\
{\rm gdmack@owu.edu, adithya.manohar.2012@owu.edu}}


\begin{abstract}
Constraints are placed on the spin-independent interaction cross section of dark matter with regular matter by refining two methods. First, dark matter--cosmic ray interactions are considered, wherein cosmic ray protons collide with dark matter to contribute to the gamma ray sky. This constraint is developed using the NFW and Moore dark matter density profiles and new data from the Fermi gamma ray space telescope. Second, the Earth capture scenario is considered, wherein particles that are captured self-annihilate at Earth's center, thus adding to its internal heat flow. The constraint presented here is developed based on analysis of the drift time of dark matter particles through Earth, modeled as a core composed of iron and a mantle composed of oxygen with linear density gradients between layers. An analysis of the cosmic ray constraint (which rules out dark matter--regular matter interaction cross sections greater than its value) shows that it overlaps significantly with the Earth drift time constraint (which rules out cross sections smaller than its value), closing the window on strongly interacting dark matter particles up to a mass of about $10^{17}$ GeV when combined with other exclusions. 
\end{abstract}



\maketitle

\section{Introduction}
\label{sec:intro}

While the evidence corroborating dark matter's existence continues to accumulate \cite{clowe} \cite{feng} \cite{berg}, the particle nature of dark matter is still unknown. It is important to formulate models of specific candidates, but it also is useful to keep the nature generic. As summarized nicely by Taoso, Bertone, and Masiero \cite{tenpoint}, dark matter at its simplest must be a neutral, cold, weakly interacting massive particle that existed in the early universe. While a weak interaction strength (corresponding to small dark matter--regular matter interaction cross sections) is strongly suggested, larger cross sections have not been completely ruled out. The parameter space of the spin-independent dark matter--nucleon interaction cross section $\sigma_{\chi N}$ and the dark matter mass $m_\chi$ continues to be whittled down, however.

Dark matter's interactions with regular matter can be grouped into three categories relative to Earth. In the first, dark matter would scatter with particles in the Galactic environment, and therefore not interact with Earth, resulting in Galactic-scale effects. In the second, dark matter would interact weakly enough to make it through the Galaxy, through Earth's atmosphere, and also through a few kilometers of Earth's surface before scattering with regular matter. Underground detectors like Xenon100 \cite{xenon}, and others like CDMS \cite{cdms} \cite{ahmed}, CRESST \cite{cresst}, and Edelweiss \cite{edel} \cite{edel2}, have ruled out areas of the dark matter mass--cross-section parameter space in this region. A third regime lies between the first two, where dark matter can pass unaffected through the Galaxy, but interacts with Earth's atmosphere or the first bit of Earth's surface. Indirect detectors like IceCube \cite{ice} and SkyLab \cite{skylab} \cite{erickcek} have helped rule out areas of the parameter space in this region.

\begin{figure}[!h] 
\includegraphics[width=3.25in,clip=true]{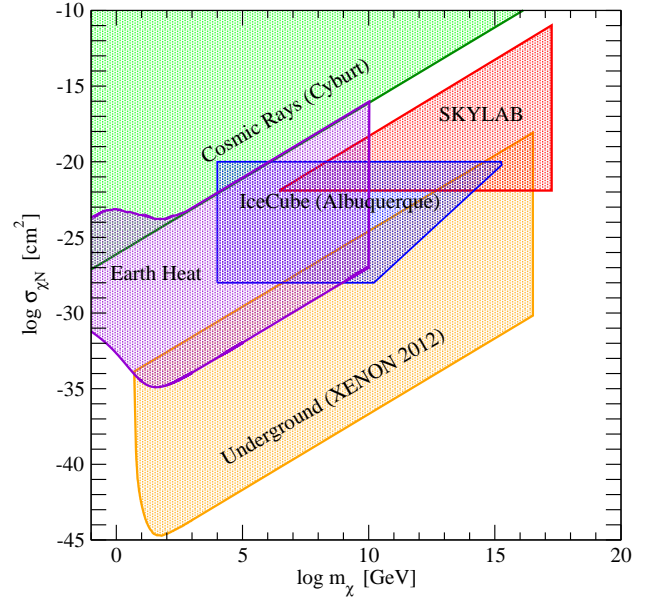}

   \caption{\label{fig:exclusion} In this log-log plot, the shaded regions of the mass--cross-section parameter space shown here are excluded (m$_{\chi}$ is the dark matter mass and $\sigma_{\chi N}$ is the dark matter--nucleon interaction cross section). Other constraints exist, but are covered by the shown regions. The values displayed are those prior to the analysis of this paper. The astrophysical limit from cosmic rays \cite{cyburt} and the latest underground detector limits (Xenon100 \cite{xenon}) are shown in green and orange, respectively. The middle regions are those from an analysis of data from SKYLAB \cite{skylab} \cite{erickcek} in red, IceCube \cite{ice} in blue, and the Earth heat scenario \cite{mbb} in purple). This paper concerns the green (upper-left) and purple (left-center) regions.}
   
\end{figure}

The findings in Mack, Beacom, and Bertone \cite{mbb} strongly suggest that dark matter must indeed be weakly interacting over a wide range of dark matter masses due to analysis of this third regime. The upper edge of the newly-constrained region from that investigation (purple region labeled ``Earth Heat" in Fig.~\ref{fig:exclusion}) lines up with the lower edge of the constraints of the first category (solid green line in the same figure), ruling out a wide swath of dark matter interaction strengths over all three regimes. However, a more forceful statement towards closing the window on strongly interacting dark matter can now be made since further research presented here has caused these constraint regions to overlap significantly. Note that in this paper ``strong" interactions refer simply to those that are significantly higher in value than ``weak" interactions.

\section{Cosmic Ray Constraints}
\label{sec:cosmic}

The Galactic contribution to the gamma ray sky comes primarily from the pion production reaction of cosmic ray protons colliding with interstellar hydrogen nuclei (protons). By analogy, collisions between cosmic ray protons ($p$) and dark matter particles ($\chi$) would also produce gamma rays of similar energy:
\begin{align*}
\chi + p \longrightarrow \chi + p + \pi^{0} \longrightarrow \chi + p + 2\gamma
\end{align*}
Following Cyburt {\it et al.}\ \cite{cyburt}, the emissivity $q_\gamma$ (number of photons per volume per time) is given by
\begin{align}
q_\gamma(r) = 2 \rho(r) \frac{\sigma_{\chi N}}{m_\chi} \Phi_p
\end{align}
where $\Phi_p$ is the angle-integrated cosmic ray proton intensity for all energies above the pion production threshold. Note that in this section, $\rho(r)$ refers to the  galactic distribution of dark matter.

This paper uses the Navarro-Frenk-White (NFW) \cite{nfw} and Moore \cite{moore} \cite{yhba} dark matter density profiles for $\rho(r)$, for $r$ in kpc. These profiles are of the form
\begin{align*}
\rho(r) = \frac{\rho_0}{(r/r_s)^\gamma [1+(r/r_s)^\alpha]^{(\beta-\gamma)/\alpha}}
\end{align*}
with the following parameters:

\begin{table}[ht]
{\small
\hfill{}
   \centering
   \caption{\label{tab:profiles} Dark matter density profiles}
   \begin{tabular}{@{} lcccccc @{}} 
      \hline
      Profile    & $\alpha$ & $\beta$ & $\gamma$ & $r_s$ (kpc) & $\rho(R_{SC})$(GeV/cm$^{3}$) & $\rho_0$(GeV/cm$^{3}$)\\
      \hline\hline
      NFW      & 1 & 3 & 1 & 20 & 0.3 & 0.259\\
      Moore    & 1.5  & 3 & 1.5 & 28 & 0.27 & 0.0527\\
      \hline
   \end{tabular}
   \hfill{}
   }
   
\end{table}
The $\rho_0$ term is chosen to ensure that the local density (at $r=R_{SC}$, the radius of the solar circle) of dark matter is 0.3 GeV cm$^{-3}$ for NFW (or 0.27 GeV cm$^{-3}$ for Moore).
Following Cyburt {\it et al.}\ \cite{cyburt}, the proton intensity $\Phi_p$ is taken to be a constant value in the high dark matter mass limit ($m_\chi \rightarrow \infty$), so that $\Phi_{p,\infty}$ = 11.8 cm$^{-2}$ s$^{-1}$. For smaller $m_\chi$, this scales as:
\begin{align}
\label{eqn:phi}
\Phi_p = \Phi_{p,\infty}\left[1+\frac{m_{\pi^0}(2m_p + m_{\pi^0})}{2m_\chi(m_p + m_{\pi^0})}\right]^{-\Gamma}
\end{align}
where $\Gamma$ $\sim$ 1.7 \cite{cyburt}. Here $\Phi_p$ represents the intensity of cosmic ray protons that are energetic enough to produce gamma rays. 


The gamma ray intensity due to these dark matter--cosmic ray interactions, viewed in any given direction, is the line integral of $q_\gamma$ along that line of sight $l$:
\begin{align}
\phi_{\gamma}= ~&{} \frac{1}{4\pi}\int_{l.o.s} q_\gamma(r)dl\\
= ~&{} \frac{1}{2\pi} \frac{\sigma}{m_\chi} \Phi_p \int_{l.o.s} \rho(r)dl.
\label{eqn:los}
\end{align}
The NFW density profile $\rho(r)$ is expressed in terms of the Galaxy's radial coordinate. Transforming this coordinate so that Earth is the origin for a line of sight through the Galactic center gives
\begin{align}
r = \sqrt{R_{SC}^2 - 2lR_{SC} + l^2}
\end{align}
where $R_{SC}$ = 8.5 kpc, the radius of the Solar Circle (or the distance from the Sun to the Galactic Center) \cite{yhba}. The NFW profile in these coordinates is
\begin{align*}
\rho(l^\prime) = \frac{\rho_0}{\left(\frac{\sqrt{72.25 - 17l^\prime + l^{\prime2}}}{20}\right)\left[1+\left(\frac{\sqrt{72.25 - 17l^\prime + l^{\prime2}}}{20}\right)\right]^2}
\end{align*}
with $l^\prime$ being $l$/kpc. 

To avoid the divergence of the integral in Eq.~\ref{eqn:los} at the Galactic Center ($l = R_{SC}$), the integral is split into three regions, with a central region of radius 0.015 kpc treated as having a uniform density $\rho_c$ of dark matter, as in Yuksel {\it et al.}\ \cite{yhba}, such that $\rho_c = \rho(8.485\mbox{ kpc}) = \rho(8.515\mbox{ kpc})$. The gamma ray intensity is then
\begin{eqnarray}
\label{eqn:integral}
\phi_{\gamma} &=& \frac{1}{2\pi} \frac{\sigma_{\chi N}}{m_\chi} \Phi_p * \nonumber \\*
&&\left[\int_0^{8.485} \rho(l^\prime)dl^\prime + \int_{8.485}^{8.515} \rho_cdl^\prime
+ \int_{8.515}^{21} \rho(l^\prime)dl^\prime \right]
\end{eqnarray}
The upper limit on the integral, 21 kpc (or a distance of 12.5 kpc from the Galactic Center along the line of sight), marks the extent of the the gamma ray region of interest; the intensity outside this region is not high enough to be significant \cite{fermi}. 

The diffuse gamma ray emission (DGE) intensity in the direction of the Galactic Center is $2.7\times10^{-4}$ cm$^{-2}$ sr$^{-1}$ s$^{-1}$ when averaged over the four models presented by Ackermann {\it et al.}\ \cite{fermi} based on the observations of Fermi-LAT. This value is further refined by accounting for contributions to the DGE by cosmic ray interactions with H{\sc i}, H{\sc ii} and H$_2$ clouds. On subtracting these contributions using an average of the models presented by Ackermann {\it et al.}, the maximum intensity attributable to dark matter interactions in the direction of the Galactic Center is $3\times10^{-5}$ cm$^{-2}$ sr$^{-1}$ s$^{-1}$. However, Ackermann {\it et al.} also show that pion production is responsible for only about two-thirds of the DGE, rather than all of it. This gives a final $\phi_\gamma$ value of $2\times10^{-5}$ cm$^{-2}$ sr$^{-1}$ s$^{-1}$. Thus, in the high dark matter mass limit, evaluating Eq.~\ref{eqn:integral} for the NFW profile gives
\begin{align}
\label{eqn:nfw}
\frac{\sigma_{\chi N}}{m_\chi} = 
~&{} 4.9 \times 10^{-29}\mbox{ cm}^2\mbox{ GeV}^{-1}.
\end{align}
Any values of  $\sigma_{\chi N}/m_\chi$ larger than this will result in too much gamma ray emission; values lower than this are allowed. If we relax the constraint that $m_\chi \rightarrow \infty$, then from Eq.~\ref{eqn:phi} we have
\begin{align}
\sigma_{\chi N} = \sigma_\infty \left[1+\frac{m_{\pi^0}(2m_p + m_{\pi^0})}{2m_\chi(m_p + m_{\pi^0})}\right]^{1.7},
\end{align}
where $\sigma_\infty$ is the cross section in the high dark matter mass limit as shown in Eq.~\ref{eqn:nfw}. This nonlinear behavior at low m$_{\chi}$ values is due to the mass dependence of $\Phi_p$ as discussed above.  This constraint (including the nonlinear behavior) is displayed in Fig.~\ref{fig:best-case}.


The Moore profile, with Earth as the origin for a line of sight through the Galactic Center, is
\begin{align*}
\rho(l^\prime) = \frac{\rho_0}{\left(\frac{\sqrt{72.25 - 17l^\prime + l^{\prime2}}}{28}\right)^{1.5}\left[1+\left(\frac{\sqrt{72.25 - 17l^\prime + l^{\prime2}}}{20}\right)^{1.5}\right]}
\end{align*}
with $l^\prime$ being $l$/kpc. Integrating to a limit of $l$ = 21 kpc in three parts (as for the NFW profile) with the same value of $\phi_\gamma$, we have
\begin{align}
\frac{\sigma_{\chi N}}{m_\chi} = ~&{} 9.3 \times 10^{-30}\mbox{ cm}^2\mbox{ GeV}^{-1},
\end{align}
which is a lower limit than that obtained for the NFW profile, as expected.

If instead the full DGE value of  $2.7\times10^{-4}$ cm$^{-2}$ sr$^{-1}$ s$^{-1}$ were used, the values for $\sigma_{\chi N}/m_\chi$  in the high dark matter mass limit become $6.6 \times 10^{-28}\mbox{ cm}^2\mbox{ GeV}^{-1}$ and $1.2 \times 10^{-28}\mbox{ cm}^2\mbox{ GeV}^{-1}$ for the NFW and Moore profiles, respectively.


\section{Earth Drift Time Constraint}


Mack, Beacom, and Bertone \cite{mbb} explored the third regime mentioned in Section \ref{sec:intro} by investigating the consequences of dark matter's interaction with regular matter---namely, dark matter's loss of energy and consequential gravitational capture by Earth. The dark matter would sink to the bottom of the gravitational well, Earth's core, where it would annihilate with other dark matter particles that experienced the same fate. (In that analysis as well as this, dark matter is assumed to be its own antiparticle.) If enough dark matter is collected through this process the subsequent annihilation would produce a heat flow coming from Earth's core that violates the measured value. This situation defines the lower edge of the exclusion region: any spin-independent dark matter--regular matter interaction cross section $\sigma_{\chi N}$ above this line would result in too much heat. (Note that in Fig.~\ref{fig:exclusion} the bottom line of the constraint from Mack, Beacom, and Bertone \cite{mbb} has been shifted downward from the original by a factor of two to reflect a calculational error in the original work.)

The upper edge of the exclusion region is where the scenario breaks down: the dark matter would interact so frequently with the regular matter that it would not make it down to the center in a sufficient timescale and therefore would not efficiently annihilate. The original analysis followed Starkman {\it et al.}\ \cite{starkman}, requiring that the dark matter be able to drift down to the bottom of the gravitational potential well in a time period of 1 Gyr. That analysis also assumed an interaction of dark matter with iron nuclei. We have refined this calculation and show that the new value is very similar to the original.

In order to more accurately represent the drifting of dark matter to Earth's center, we break its interior into layers, each with a corresponding size, density, and dominant target nucleus for interaction. Following Mack, Beacom, and Bertone, to prevent the unlikely consequence that any dark matter particle merely coming into contact with Earth will be captured (and to exclude glancing trajectories, for instance), only those particles that travel a path length of at least 0.2 R$_\oplus$ are considered to be captured \cite{mbb}. Thus, the effective scattering volume for Earth has a radius of 0.99 R$_\oplus$, or roughly R = 6300 km.

Four layers of the Earth were considered in order to accurately model dark matter interactions in its interior: the inner core (i), the outer core (o), the lower mantle (l), and the upper mantle (u), as shown in Table 1. This analysis treats the core layers as composed entirely of iron, and the mantle layers as composed entirely of oxygen. The density distribution of each layer was treated as a linear gradient between the values at the layer boundaries. Density values are approximated from the Preliminary Reference Earth Model (PREM) \cite{prem}, and temperature values are conservative estimates of average layer temperatures. Each gradient produced a layer mass consistent with PREM.
\begin{table}[ht]
{\small
\hfill{}
     \caption{  \label{tab:model} Earth model setup} 

   \begin{tabular}{@{} lccc @{}} 
      \hline
      Layer    & Height (km) & Density (g/cm$^{3}$) & Temp (K)\\
     \hline\hline
      Inner Core (i)      & 0 - 1200 & 13.1 - 12.8 & 6000 \\
      Outer Core (o)      & 1200 - 3500  & 12.2 - 9.9 & 5000 \\
      Lower Mantle (l)       & 3500 - 5700  & 5.6 - 4.4 & 3500\\
      Upper Mantle (u) & 5700 - 6300   &  4.0 - 3.4 & 1500 \\
      \hline
   \end{tabular}
   \hfill{}}
 
\end{table}

For a dark matter particle in equilibrium, with a constant drift velocity, the gravitational force on it due to Earth equals the drag force on it:
\begin{align}
\label{eqn:balance}
\frac{GM(r)m_\chi}{r^2} = \sigma_{\chi A} v_{th}nm_{p}v_{d}
\end{align}
where G is the universal gravitational constant, $M(r)$ is the mass of the earth interior to the particle at radius $r$, $m_\chi$ is the mass of the dark matter particle, $\sigma_{\chi A}$ is the interaction cross section of dark matter with target particles of mass number $A$, $v_\mathrm{th}$ is the thermal speed of the target particles, $n$ is the number density of the target particles, $m_A$ is the mass of the target particle, and $v_{d}$ is the drift velocity of the dark matter particle. 

The drift velocity can then be expressed as distance traveled over drift time ($v_{d} = r/\tau_{d}$), and the mass of the target multiplied by the number density can be expressed as the density of target particles ($nm_A = \rho(r)$), while the thermal velocity of the target particles can be expressed as that of an ideal gas ($v_\mathrm{th} = \sqrt{6kT/m_A}$) 
with the temperature being conservatively estimated as an average for each layer. Eq.~\ref{eqn:balance} can then be expressed in terms of the drift time.

Note that $\rho$ in this section refers to the density of the target particles in Earth. Since $\rho$ is a function of radius, a small time interval $d\tau_d$ in which the particle drifts a distance $dr$ is
\begin{align}
\label{eqn:dt}
{d\tau_{d}} = \left(\frac{\sigma_{\chi A}}{m_\chi}\right) \frac{v_\mathrm{th}}{G}\times\frac{d}{dr}\left(\frac{r^3\rho(r)}{M(r)}\right)dr,
\end{align}
where the density function $\rho(r)$ for each layer was derived by constructing a linear gradient between the layer boundaries.

Integrating Eq.~\ref{eqn:dt} between the layer boundaries using the density gradients gives a value for the drift time through a specific layer, in terms of $\sigma_{\chi A}/m_\chi$. The integral was taken until a radius of 100 m, which marks the effective radius of a collected volume of dark matter, where the particle can annihilate \cite{mbb} \cite{greist}. For the timescale imposed, the total of all four $\tau$ values must be 1 Gyr: 
\begin{eqnarray}
1\mbox{ Gyr} &=&  \tau_i + \tau_o + \tau_l + \tau_u\\
&=&  \frac{\sigma_{\chi Fe}}{m_\chi}(7.66\times10^{12}\mbox{ kg s m}^{-2})\nonumber \\
  &~& + \frac{\sigma_{\chi O}}{m_\chi}(3.76\times10^{12}\mbox{ kg s m}^{-2})
\label{eqn:drift}
\end{eqnarray}
This more usefully should be put in terms of $\sigma_{\chi N}/m_\chi$, which is the interaction cross section of dark matter with a single nucleon, rather than a specific target:
\begin{align}
\label{eqn:red}
\sigma_{\chi A} = ~&{} \sigma_{\chi N} A^{2}\left(\frac{\mu_A}{\mu_N}\right)^2\\
= ~&{} \sigma_{\chi N} A^{2}\left(\frac{m_\chi m_A}{m_\chi+m_A}\right)^2 \left(\frac{m_\chi+m_N}{m_\chi m_N}\right)^2\\
= ~&{} \sigma_{\chi N} A^{4}\left(\frac{m_\chi+m_N}{m_\chi+m_A}\right)^{2}
\label{eqn:correction}
\end{align}
where $\mu$ is the reduced mass of the dark matter particle and the target, which is either a nucleus ($A$) or a nucleon ($N$), and $A$ is the mass number of the target particle. The $A^{2}$ term indicates that at these low momentum transfers, the target nucleus is not resolved, and it is assumed that the dark matter particle couples coherently to the net ``charge" or the number of nucleons \cite{mbb}. Note that in the high dark matter mass limit ($m_\chi >> m_A$), the enhancement term reduces to $A^4$.

\begin{figure}[!ht] 
   \centering 
   \includegraphics[width=3.25in, clip=true]{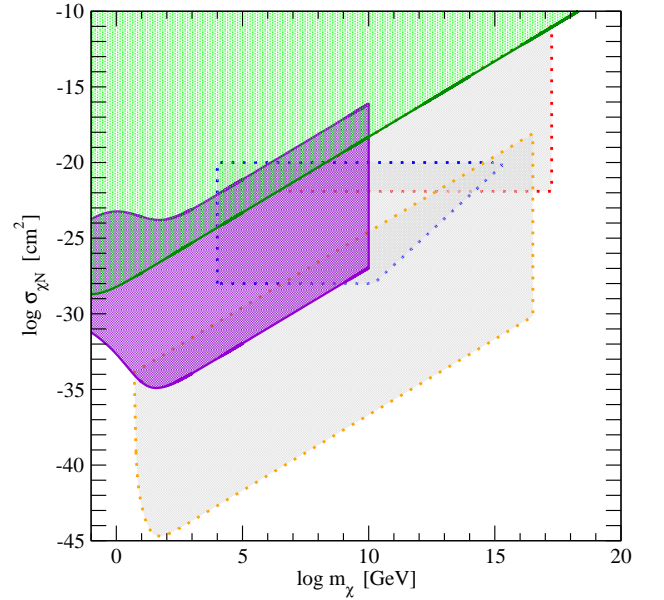}
   \caption{\label{fig:best-case} In this log-log plot, updated exclusion regions as determined in this paper are shown in green (upper-left) and purple (left-center) with solid border lines. Shaded regions are excluded. The cosmic ray constraint shown here (upper left) is based on the NFW profile, but it accounts for contributions to the DGE by interstellar gas clouds and the pion production contribution. Note that using the Moore profile would lower this constraint by a factor of $\sim$5, increasing the overlap. The cosmic ray constraint not only overlaps with the Earth drift time constraint, but also rules out large cross sections for dark matter masses up to $\sim$10$^{17}$GeV when combined with the SkyLab, IceCube, and underground detector constraints.} 
\end{figure}

From Eq.~\ref{eqn:drift} and Eq.~\ref{eqn:correction}, for $m_\chi >> m_A$  using $A=56$ for Fe and $A=16$ for O this becomes
\begin{align}
\frac{\sigma_{\chi N}}{m_\chi} = 7.4 \times 10^{-27}\mbox{ cm}^{2}\mbox{ GeV}^{-1}.
\end{align} 
Values of $\sigma_{\chi N}/m_\chi$ lower than this result but greater than the bottom edge of the Earth heat constraint in Ref. \cite{mbb} are excluded; values greater than this result are allowed. The behavior of $\sigma_{\chi N}$ when the high dark matter mass limit is relaxed is shown in Fig.~\ref{fig:best-case}. The numerical value varies insignificantly from the original  result of  $7.8 \times10^{-27}$ cm$^{2}$ $m_\chi$/GeV, but the behavior at low masses reflects the addition of oxygen as a target of interest. This very slightly alters the curvature of the limit around $m_{\chi}$ = 16 GeV, which reflects the fact that the most efficient scattering occurs when the dark matter particle mass equals the target nucleus mass. The insignificant variance of this result from the previous version shows the robustness of the calculation. Note that this value is more than two orders of magnitude above the NFW profile limit obtained for the cosmic ray-dark matter interactions in the previous section. 

\section{Conclusions}

For the Earth capture and drift time analysis, a constraint of $\sigma_{\chi N}$ =  $7.4 \times10^{-27}$ cm$^{2}$ $m_\chi$/GeV was obtained (in the high dark matter mass limit) as the upper limit of all cross sections that would lead to Earth overheating due to dark matter accumulation and annihilation. By comparison, the original Earth drift time constraint obtained by Mack, Beacom, and Bertone \cite{mbb} was $7.8 \times10^{-27}$ cm$^{2}$ $m_\chi$/GeV in the high dark matter mass limit. The insignificant variance in the results shows the robustness of the situation and calculation.

Based on Fermi-LAT observations of the gamma ray sky, taking into account the contributions to the DGE by cosmic ray interactions with H{\sc i}, H{\sc ii}, and H$_2$ and the fraction of the gamma rays produced by pion decay, a constraint of $\sigma_{\chi N}$ = $4.9 \times10^{-29}$ cm$^{2}$ $m_\chi$/GeV (in the high dark matter mass limit) was obtained for the NFW profile; cross sections larger than this value are excluded. The Moore profile provides a stronger overlap, as expected, yielding a value of $9.3 \times10^{-30}$ cm$^{2}$ $m_\chi$/GeV for the same high dark matter mass limit. By comparison, the original cosmic ray interaction constraint obtained by Cyburt {\it et al.}\ \cite{cyburt} was $4.6 \times10^{-27}$ cm$^{2}$ $m_\chi$/GeV in the high dark matter mass limit, using an isothermal dark matter profile.

As shown in Fig.~\ref{fig:best-case}, the NFW constraint (which is more conservative than the Moore constraint) is more than two orders of magnitude smaller than the  upper edge of the Earth drift time constraint. This is a strong statement excluding all cross sections in that region, definitively ruling out strongly interacting dark matter up to a mass of $10^{10}$ GeV (since the Earth capture scenario cannot be applied above this mass \cite{mbb}). When combined with the SkyLab, IceCube, and underground detector constraints, these results serve to rule out strongly interacting dark matter up to a  mass of $\sim10^{17}$ GeV.  Note that if the more conservative approach of using the full value of the DGE were taken for comparison, the cosmic ray limit still overlaps the Earth drift time limit by an order of magnitude, ruling out strongly interacting dark matter up to a mass of 10$^{10}$ GeV. The results in this paper show more definitively that the dark matter--nucleon spin-independent interaction cross section must indeed be weak for a very large range of dark matter masses. Further analysis can be done looking away from the Galactic Center, where the gamma ray emission falls much more sharply than the dark matter distribution.

\vspace{0.4in}

We wish to thank John Beacom of The Ohio State University and CCAPP for his helpful comments, and the Department of Physics and Astronomy at Ohio Wesleyan University for its support.

\end{document}